\documentclass[11pt]{article}
\usepackage{cite}
\usepackage{amsmath,amsfonts,amssymb}
\usepackage{pstricks,pst-node}
\usepackage[small,bf,hang]{caption2}

\psset{unit=1.2cm,linewidth=.5pt,radius=.2}

\newcommand{\normalNode}[2]{\Cnode(#1){#2}}
\newcommand{\dualityNode}[2]{\Cnode[fillstyle=solid,fillcolor=lightgray](#1){#2}}
\newcommand{\disabledNode}[2]{\Cnode[fillstyle=solid,fillcolor=black](#1){#2}}
\newcommand{\singleConnection}[2]{\ncline{-}{#1}{#2}}

\def\hybrid{
        \topmargin -20pt
        \oddsidemargin 0pt
        \headheight 0pt \headsep 0pt
        \textwidth 6.25in       
        \textheight 9.5in       
        \marginparwidth .875in
        \parskip 5pt plus 1pt   \jot = 1.5ex}

\hybrid

\newcommand{\cA}{{\cal A}}
\newcommand{\cB}{{\cal B}}

\newcommand{\cK}{{\cal K}}
\newcommand{\cL}{{\cal L}}
\newcommand{\cM}{{\cal M}}
\newcommand{\cN}{{\cal N}}

\newcommand{\cP}{{\cal P}}
\newcommand{\cQ}{{\cal Q}}
\newcommand{\cR}{{\cal R}}
\newcommand{\cS}{{\cal S}}

\newcommand{\cV}{{\cal V}}

\newcommand{\beq}{\begin{equation}}
\newcommand{\eeq}{\end{equation}}
\newcommand{\bi}{\begin{itemize}}
\newcommand{\ei}{\end{itemize}}
\newcommand{\bea}{\begin{eqnarray}}
\newcommand{\eea}{\end{eqnarray}}
\newcommand{\ba}{\begin{array}}
\newcommand{\ea}{\end{array}}
\newcommand{\bt}{\begin{tabular}}
\newcommand{\et}{\end{tabular}}
\newcommand{\bc}{\begin{center}}
\newcommand{\ec}{\end{center}}

\newcommand{\ft}[2]{{\textstyle {\frac{#1}{#2}} }}

\newcommand{\pf}[1]{\boldsymbol{\mathit{#1}}}
\newcommand{\R}{\mathbb{R}}

\begin{document}

\begin{titlepage}
\begin{center}

\hfill UG-08-04 \\

\vskip 1.5cm

{\LARGE \bf A Note on $E_{11}$ and \\Three-dimensional  Gauged
Supergravity
\\[0.2cm]}

\vskip 1.5cm

{\bf Eric A.~Bergshoeff, Olaf Hohm, and Teake A. Nutma} \\

\vskip 25pt

{\em Centre for Theoretical Physics, University of Groningen, \\
Nijenborgh 4, 9747 AG Groningen, The Netherlands \vskip 5pt }

{email: {\tt E.A.Bergshoeff@rug.nl, O.Hohm@rug.nl, T.A.Nutma@rug.nl}} \\

\vskip 0.8cm

\end{center}

\vskip 1cm

\begin{center} {\bf ABSTRACT}\\[3ex]

\begin{minipage}{13cm}
We determine the gauge symmetries of all $p$--forms in maximal
three-dimensional gauged supergravity ($0\leq p\leq 3$) by requiring
invariance of the Lagrangian. It is shown that in a particular
ungauged limit these symmetries are in precise correspondence to
those predicted by the very-extended Kac-Moody algebra $E_{11}$. We
demonstrate that whereas in the ungauged limit the bosonic gauge
algebra closes off-shell, the closure is only on-shell in the full
gauged theory. This underlines the importance of dynamics for
understanding the Kac-Moody origin of the symmetries of gauged
supergravity.

\end{minipage}

\end{center}

\noindent

\vfill

March 2008

\end{titlepage}

\section{Introduction}\setcounter{equation}{0}

One of the surprising results of supergravity is that the
Kaluza-Klein reduction of the maximal 11-dimensional theory on a
$d$-torus yields the exceptional hidden symmetry groups $E_{d(d)}$
for $6\leq d\leq 9$ \cite{Julia:1982gx}. This has led to the
conjecture that the over-extended and very-extended Kac-Moody
algebras $E_{10}$ \cite{Nicolai:2003fw, Damour:2002cu,
Damour:2004zy,Kleinschmidt:2004rg} and $E_{11}$
\cite{West:2001as,Schnakenburg:2001ya,West:2002jj,Kleinschmidt:2003mf,West:2005gu,Englert:2003py,Englert:2003zs}
may be of relevance for the original theory or, more optimistically,
be even the ultimate symmetry of M-theory.

Recently, it has been shown that $E_{11}$ (and to some extend also
$E_{10}$) contains information about the possible deformations of
supergravity into gauged or massive supergravities
\cite{Riccioni:2007au,
Bergshoeff:2007qi,Bergshoeff:2007vb,Riccioni:2008jz}. More
precisely, a level decomposition shows that the spectra of $E_{11}$
and $E_{10}$ contain $(D-1)$-form  potentials that, via duality, are
in precise correspondence with the embedding tensor $\Theta$
introduced in \cite{Nicolai:2000sc,Nicolai:2001sv} for maximal
gauged supergravity in $D=3$ (and subsequently generalized to higher
dimensions in \cite{deWit:2002vt, deWit:2003ja, deWit:2003hr,
deWit:2004nw, deWit:2005hv,
 Samtleben:2005bp,deWit:2007mt,Bergshoeff:2007ef,deWit:2008ta}).
In addition, the spectrum of $E_{11}$ contains $D$-form potentials
that are in part related to quadratic constraints on the embedding
tensor \cite{Bergshoeff:2007vb,deWit:2008ta}.

The embedding tensor approach is based on the introduction of a
tensor $\Theta$ that is in a particular representation of the
 duality group and which encodes the gauging.
 A special feature of the three-dimensional maximally
 supersymmetric case is
 that all bosonic matter fields can be dualized to scalars leading
 to a 128-dimensional $E_{8(8)}/SO(16)$ coset space. However, to
 gauge a subgroup of the duality group one needs to introduce
 vectors as well. It was shown in \cite{Nicolai:2000sc,Nicolai:2001sv}
 that this can be  achieved by a topological term of the
 form $\Theta A\partial A+ \Theta^2 A^3$, where $A$
 are the gauge vectors, which in turn do not lead to new degrees
 of freedom.
In higher dimensions, a whole hierarchy of $p$--form potentials with
$0 \le p \le D-2$ is introduced \cite{deWit:2005hv,deWit:2008ta}. It
is a generic feature of this hierarchy that the gauge algebra can be
closed off-shell.

 For consistency the embedding tensor has to satisfy
 a set of quadratic constraints.
 Given a gauged supergravity theory containing the constant
 embedding tensor one can promote this tensor to an unconstrained scalar
 field
 $\Theta(x)$ by adding to the original Lagrangian ${\cal L}_{g}$
 a further topological term containing the deformation and top-form potentials
 as Lagrange multipliers in the following way
 \cite{Bergshoeff:2007vb,deWit:2008ta}:
 \begin{equation}\label{action1}
 {\cal L} = {\cal L}_{g} + A_{(D-1)}\partial\Theta +
 A_{(D)}\Theta\Theta\,,
 \end{equation}
where we have suppressed the duality and space-time indices. These
extra potentials complete the hierarchy of potentials to include all
$p$--forms with $0\le p \le D$. There is, however, a subtlety with
the bosonic gauge transformations of these new potentials. The
gauge-invariance of the original Lagrangian ${\cal L}_{g}$ will
be violated by terms proportional to either $\partial\Theta$ or
$\Theta^2$. Such terms can always be be canceled by assigning
bosonic gauge transformations to the deformation and top-form
potentials. However, it is not obvious that the gauge
transformations determined like this coincide with those derived
from the general formalism, which is valid for the full hierarchy of
$p$--forms in generic dimension. In fact, by inspecting closure of
the supersymmetry algebra it has already been pointed out in
\cite{deWit:2008ta} that the gauge transformations receive
modifications when applied to a specific model. Here we are going to
derive the full bosonic gauge symmetries for three-dimensional
gauged maximal supergravity directly by requiring invariance of the
Lagrangian \eqref{action1}. In particular, we will find that the
closure is only on-shell.

Moreover, we are going to compare the resulting symmetries with
those predicted by $E_{11}$. Since the latter does not give rise to
the embedding tensor, but only to its dual deformation potential,
naively this would require to take the ungauged limit, i.e.~to set
the embedding tensor equal to zero.\footnote{Recently, a scheme has
been proposed to include the embedding tensor via a further
extension of $E_{11}$ \cite{Riccioni:2007ni}. Here we will not
explore this possibility.} However, we will see that in this limit
terms survive in the transformation rules that are not predicted by
$E_{11}$. Instead, we will define a different limit, in which the
symmetries precisely match and which, moreover, has the advantage
that all $p$--forms but the top-form survive in the action. We will
also see that in this limit the bosonic gauge algebra reduces to an
algebra that closes off-shell, in accordance with the level
decomposition of $E_{11}$.


This note is organized as follows. In Section 2 we first introduce
the maximal gauged supergravity theory in three dimensions,
following \cite{Nicolai:2000sc,Nicolai:2001sv,deWit:2008ta}. Then we
give the complete bosonic gauge transformations of all $p$--form
potentials and show that the bosonic gauge algebra closes on-shell.
In the next section we perform the level decomposition of $E_{11}$
and show how the result obtained agrees with a particular limit of
the gauged supergravity result discussed in Section 2. In this limit
the on-shell closed gauge algebra reduces to an off-shell closed
one. Finally, in the conclusions we comment about the consequences
of our results for a Kac-Moody approach to gauged supergravity in
general.

\section{Gauged supergravity in $D=3$}
In this section we  give a brief review of gauged maximal
supergravity in $D=3$ \cite{Nicolai:2000sc,
Nicolai:2001sv,deWit:2008ta}. In the first subsection we will
introduce the Lagrangian and the embedding tensor.
In the following subsection
we will introduce an equivalent formulation \cite{deWit:2008ta}, in
which the non-propagating 2-form and 3-form fields predicted by
$E_{11}$ appear, and determine their bosonic gauge symmetries.

\subsection{The Lagrangian and the embedding tensor}
The propagating bosonic degrees of freedom of maximal supergravity
in $D=3$ consist of 128 scalar fields parameterizing the coset space
$E_{8(8)}/SO(16)$. Besides, there are the topological metric and, in
gauged supergravity, Chern-Simons vectors. The 128 scalars are
encoded in the $E_{8(8)}$ valued matrix ${\cal V}^{\cM}{}_{\cA}$,
where $\cM,\cA,\ldots =1,\ldots,248$ denote adjoint indices of
$E_{8(8)}$. We indicate by letters from the middle and the beginning
of the alphabet `curved' indices corresponding the global left
action and `flat' indices corresponding to the local right action,
respectively. The scalars enter the Lagrangian via the non-compact
part of the Maurer-Cartan forms
 \bea
  {\cal V}^{-1}D_{\mu}{\cal V} \ = \ \ft12
  Q_{\mu}^{IJ}X^{IJ}+P_{\mu}^{A}Y^{A}\;,
 \eea
which we wrote according to the $SO(16)$ decomposition ${\bf
248}={\bf 120}\oplus {\bf 128}$. Here $X^{IJ}$ denote the $SO(16)$
generators, with vector indices $I,J,\ldots = 1,\ldots,16$, and
$Y^{A}$ are the non-compact generators transforming as spinors under
$SO(16)$, i.e.~with spinor indices $A,B,\ldots =
1,\ldots,128$.\footnote{Our $E_{8(8)}$ conventions are as in
\cite{Nicolai:2001sv}. For other decompositions of $E_{8(8)}$ and
their application to maximal gauged supergravity see
\cite{Fischbacher:2002fx,Fischbacher:2003yw,Hohm:2005ui}.}

In order for the Maurer-Cartan forms to be invariant under the local
transformations
 \bea\label{gaugerig}
  \delta{\cal V} \ = \ \hat{g}(x){\cal V}\;, \qquad
  \hat{g}\in\frak{g}_0\subset\frak{e}_{8(8)}\;,
 \eea
we introduced a gauge-covariant derivative,
 \bea
  {\cal V}^{-1}D_{\mu}{\cal V} \ = \ {\cal V}^{-1}\partial_{\mu}{\cal
  V} - gA_{\mu}{}^{\cM}\Theta_{\cM\cN}({\cal V}^{-1} t^{\cN} {\cal V})\;,
 \eea
where $g$ is the gauge coupling constant. The symmetric tensor
$\Theta_{\cM\cN}$ is the embedding tensor, which encodes the
embedding of the gauge group $G_0$ into the global symmetry group
$E_{8(8)}$. More precisely, the gauge algebra $\frak{g}_0$ is
spanned by
 \bea\label{gaugepro}
   X_{\cM} \ = \ \Theta_{\cM\cN}t^{\cN}\;,
 \eea
in which $t^{\cM}$ denote the global $\frak{e}_{8(8)}$ symmetry
generators with structure constants $f^{\cM\cN}{}_{\cK}$. In
particular, the dimension of $\frak{g}_0$ is given by the rank of
$\Theta_{\cM\cN}$. In this formalism, the gauging takes a fully
$E_{8(8)}$ covariant form, since all indices are $E_{8(8)}$ indices.
Nevertheless, the duality group is no longer a symmetry due to the
fact that the constant $\Theta$ cannot transform under $E_{8(8)}$.
Rather, it acts as a projector, which breaks the symmetry down to
the gauge group $G_0$ in \eqref{gaugepro}.\footnote{Alternatively,
one could say that $E_{8(8)}$ transforms one theory into another
theory with different values of the constant $\Theta$.}

The gauged supergravity is described by the Lagrangian
 \bea\label{gaugedsugra}
 \begin{split}
  {\cal L}_g \ = \ &-\ft14 eR + \ft14 e P^{\mu A}P_{\mu
  A} - eV \\
  &-\ft14g\varepsilon^{\mu\nu\rho}A_{\mu}{}^{\cM}\Theta_{\cM\cN}(\partial_{\nu}A_{\rho}{}^{\cN}
  -\ft13 g\Theta_{\cK\cS}f^{\cN\cS}{}_{\cL} A_{\nu}{}^{\cK}
  A_{\rho}{}^{\cL})\;,
 \end{split}
 \eea
where we ignored the fermionic terms. The scalar potential $V$ is
completely determined by $\Theta$ via the so-called
T-tensor,\footnote{Following \cite{deWit:2008ta} we use a vertical
bar to distinguish between the two indices of $T$.}
 \bea
  T_{\cA|\cB} \ = \ {\cal V}^{\cM}{}_{\cA}{\cal V}^{\cN}{}_{\cal
  B}\Theta_{\cM\cN}\;.
 \eea
 Explicitly, one has
 \bea
  V \ = \
  -\frac{1}{8}g^2\left(A_1^{IJ}A_1^{IJ}-\ft12A_2^{I\dot{A}}A_2^{I\dot{A}}\right)\;,
 \eea
where
 \bea\label{a1}
  A_1^{IJ} \ = \
  \frac{8}{7}\theta\delta^{IJ}+\frac{1}{7}T_{IK|JK}\;, \qquad
  A_2^{I\dot{A}} \ = \ -\frac{1}{7}\Gamma^J_{A\dot{A}}T_{IJ|A}\;.
 \eea
 Here $\theta \equiv \tfrac{1}{248}\eta^{\cM\cN}\Theta_{\cM\cN}=\tfrac{1}{248} \eta^{\cA\cB}T_{\cA|\cB}$
 with the Cartan-Killing metric $\eta^{\cM\cN}$.
 The particular combinations $A_1$ and $A_2$ in (\ref{a1}) also
 enter  the supersymmetry variations of the fermions \cite{Nicolai:2001sv}.
In the following we give a reformulation of the scalar potential in
terms of the $E_{8(8)}$ matrix $G^{\cM\cN}=\cV^{\cM}{}_{\cA}{\cal
V}^{\cN}{}_{\cB}\delta^{\cA\cB}$. Using the inverse of the relations
\eqref{a1} \cite{Nicolai:2001sv}, we find\footnote{For performing
the required gamma matrix calculations we used the Mathematica
package GAMMA \cite{Gran:2001yh}.}
\begin{align}
V & = \tfrac{1}{32} g^2G^{\cM\cN,\cK\cL} \Theta_{\cM\cN}
\Theta_{\cK\cL} \;,
\end{align}
where
 \bea
  G^{\cM\cN,\cK\cL} \ = \ \tfrac{1}{14} G^{\cM\cK} G^{\cN\cL}
        + G^{\cM\cK} \eta^{\cN\cL}
        - \tfrac{3}{14} \eta^{\cM\cK} \eta^{\cN\cL}
        -\tfrac{4}{6727}\eta^{\cM\cN}\eta^{\cK\cL} \;.
 \eea

Note that the Chern-Simons term in \eqref{gaugedsugra} has the
effect that varying with respect to the gauge fields
$A_{\mu}{}^{\cM}$ one obtains a duality relation between the vectors
and scalars,
 \bea\label{covdual}
  e^{-1}\varepsilon^{\mu\nu\rho}\Theta_{\cM\cN}F_{\nu\rho}{}^{\cN} \ = \
  -2\Theta_{\cM\cN}{\cal V}^{\cN}{}_AP^{\mu A} \ \equiv \
  -2\Theta_{\cM\cN}J^{\mu \cN}\;.
 \eea
Here we introduced the current $J_{\mu}{}^{\cM}$, which in the
ungauged theory is the Noether current corresponding to the global
$E_{8(8)}$ symmetry. However, in the gauged theory this symmetry is
broken, and therefore the covariant conservation is violated by
terms of order ${\cal O}(g)$ induced by the scalar potential,
 \bea
  D_{\mu}\left( eJ^{\mu \cM}\right) \ = \ {\cal O}(g)\; .
 \eea
We emphasize that \eqref{covdual} is not the `naive' duality
relation in that both sides appear projected by the embedding
tensor. Consequently, only those vector fields participating in the
gauging enter \eqref{covdual}, which therefore cannot be used to
eliminate the full 248 vector fields in terms of the scalars. As has
been noted in \cite{deWit:2008ta}, there is one `extra' gauge
symmetry related to the duality relation,
 \bea\label{extra}
  \delta_{\chi} A_{\mu}{}^{\cM} \ = \
  \xi^{\nu}_{\chi}\big(F_{\mu\nu}{}^{
  \cM}+\tilde{J}_{\mu\nu}{}^{\cM}\big)\;,
 \eea
where we defined the Hodge dual
$\tilde{J}_{\mu\nu}{}^{\cM}=e\varepsilon_{\mu\nu\rho}J^{\rho \cM}$
of the current in \eqref{covdual}. Due to the missing contraction
with $\Theta_{\cM\cN}$, this is not an equations-of-motion symmetry,
but nevertheless leaves the action invariant. Though \eqref{extra}
seems to be necessary for closure of the supersymmetry algebra
\cite{deWit:2008ta}, we will not encounter this symmetry any further
in this paper.

The Lagrangian \eqref{gaugedsugra} is invariant under the gauge
transformations \eqref{gaugerig} and the following gauge
transformations of the vector potentials
 \bea\label{gaugetrans}
  \delta A_{\mu}{}^{\cM} \ = \ D_{\mu}\Lambda^{\cM} \ \equiv \ \partial_{\mu}\Lambda^{\cM} -
  gf^{\cM\cN}{}_{\cK}\Theta_{\cN\cL}A_{\mu}{}^{\cL}\Lambda^{\cK}\;,
 \eea
where the gauge parameter is related to the transformation
\eqref{gaugerig} via $\hat{g}=g\Lambda^{\cM}\Theta_{\cM\cN}t^{\cN}$.
Even though \eqref{gaugetrans} seems to describe a $248$-dimensional
local symmetry, it is actually more subtle, since the gauge vectors
$A_{\mu}{}^{\cM}$ and their variations appear in the Lagrangian
always contracted with the embedding tensor, which in turn reduces
the number of independent vector fields to ${\rm
dim}\hspace{0.1em}G_0={\rm rank}(\Theta)$. Moreover, the embedding
tensor has to satisfy a number of constraints in order for the
action to be invariant under the various symmetries. First of all,
consistency with local supersymmetry implies a linear constraint on
$\Theta_{\cM\cN}$: a priori it takes values in the symmetric tensor
product
 \bea
  ({\bf 248}\otimes {\bf 248})_{\rm sym} \ = \ \underline{{\bf 1}}\oplus \underline{{\bf
  3875}}\oplus {\bf 27000}\;,
 \eea
but supersymmetry requires that only the underlined representations
appear. Note that the singlet component of the embedding tensor
corresponds to a gauging of the full $E_{8(8)}$ duality group. In
the following we will denote symmetrization in two adjoint indices
$\cM, \cN$ and subsequent projecting away the ${\bf 27000}$
representation by $\langle\cM\cN\rangle$, e.g.
\begin{equation}
\Theta_{\cM\cN} = \Theta_{\langle\cM\cN\rangle}\,,
\end{equation}
where the explicit form of the projector has been determined in
\cite{Koepsell:1999uj}.

Secondly, invariance of the embedding tensor (and thus gauge
invariance of the action \eqref{gaugedsugra} under
\eqref{gaugetrans}), requires the quadratic constraint
\cite{Nicolai:2001sv}
 \bea\label{quadconstr}
  \cQ_{\cM\cN,\cP} \ \equiv \ \Theta_{\cK\cP}\Theta_{\cL(\cM}f^{\cK\cL}{}_{\cN)} \ = \
  0\;.
 \eea
From this definition one infers that the quadratic constraint
satisfies
\bea\label{quadconstrconstr}
  \cQ_{(\cM\cN,\cP)} \ = \ 0\;, \qquad \eta^{\cM\cN}\cQ_{\cM\cN,\cP}
  \ = \ 0\;.
  \eea
Note that for $GL(n)$ groups the first condition would imply that
$\cQ$ lives in an irreducible representation.\footnote{The projector
which implements this condition 
reads
$X^{\langle\cM\cN,\cP\rangle}=\ft23(X^{(\cM\cN),\cP}-X^{\cP(\cM,\cN)})$.}
However, this does not hold for $E_{8(8)}$, and the representation
content of (\ref{quadconstr}) can be analyzed as follows
\cite{deWit:2008ta}. Due to the linear constraint on $\Theta$, the
symmetric indices of $\cQ_{\cM\cN,\cP}$ will be in $\bf{3875}$,
where the absence of the singlet follows by the second equation in
\eqref{quadconstrconstr}. Naively the quadratic constraint
(\ref{quadconstr}) takes therefore values in
 \bea\label{quadrep}
  \bf{3875}\otimes \bf{248} \ = \ \bf{248}\oplus \underline{\bf{3875}}\oplus
  \bf{30380}\oplus\underline{\bf{147250}}\oplus\bf{779247}\;.
 \eea
However, the first condition in (\ref{quadconstrconstr}) implies
that all representations contained in the totally symmetric tensor
product $(\bf{248}\otimes \bf{248}\otimes \bf{248})_{\rm sym}$ will
be absent. This in turn reduces the irreducible representations of
$\cQ_{\cM\cN,\cP}$ to those underlined in (\ref{quadrep}). By abuse
of notation we will denote the projector onto these representations
also by brackets $\langle\hspace{0.3em}\rangle$, but note that its
explicit form is not required for our analysis.

\subsection{Deformation and top-form potentials}

We will now present an equivalent reformulation of the gauged
supergravity Lagrangian \eqref{gaugedsugra}, in which so-called
deformation and top-form potentials appear. This turns out to be
necessary in order to match the spectrum predicted by $E_{11}$.
Formally, this can be understood as follows. As we noted above, the
gauged supergravity is not invariant under $E_{8(8)}$, since as
`coupling constants', the $\Theta_{\cM\cN}$ do not transform under
the duality group. Promoting the embedding tensor to a dynamical,
i.e.~space-time dependent field $\Theta_{\cM\cN}(x)$, such that it
transforms under global rotations according to its index structure,
gives back the full $E_{8(8)}$ invariance. However, this violates
the supersymmetry and gauge invariance by terms proportional to
$\partial_{\mu}\Theta_{\cM\cN}$. This can be compensated by adding a
2-form potential to the action, and by assigning appropriate
supersymmetry and gauge variations to it. Moreover, the quadratic
constraint \eqref{quadconstr} on $\Theta_{\cM\cN}$ can be
implemented on-shell by means of a Lagrange multiplier term
containing a top-form (3-form) potential. In total we extend the
action to \cite{deWit:2008ta,Bergshoeff:2007vb}
 \bea\label{defaction}
  {\cal L}_{\rm tot} \ = \ {\cal L}_g+
  \ft14 g\varepsilon^{\mu\nu\rho}D_{\mu}\Theta_{\cM\cN}
  B_{\nu\rho}{}^{\cM\cN} -
  \ft16g^2\Theta_{\cK\cP}\Theta_{\cL(\cM}f^{\cK\cL}{}_{\cN)}
  \varepsilon^{\mu\nu\rho}C_{\mu\nu\rho}{}^{\cM\cN,\cP}\;,
 \eea
where the embedding tensor now satisfies only the linear constraint.
Consequently, the deformation potential takes values in
${\bf 1}\oplus {\bf 3875}$, while the top-form lives in ${\bf
3875}\oplus {\bf 147250}$, in accordance with (\ref{quadrep}). We
have defined a formal covariant derivative $D_\mu\Theta_{\cM\cN}$ as
 \bea\label{covder}
  D_{\mu}\Theta_{\cM\cN} \ = \ \partial_{\mu}\Theta_{\cM\cN}
  -2gA_{\mu}{}^{\cP}\Theta_{\cK\cP}\Theta_{\cL(\cM}f^{\cK\cL}{}_{\cN)}\;.
 \eea
The combination $D_\mu\Theta_{\cM\cN}$ is strictly speaking not a
covariant derivative. It would be the covariant derivative if
$\Theta_{\cM\cN}$ would transform under the gauge group according to
its index structure. However, it is convenient to set up the
calculation using a basis of gauge transformations in which the
embedding tensor is gauge-invariant,
$\delta_\Lambda\Theta_{\cM\cN}=0$. This can always be achieved by
redefining the gauge transformations with an extra equation of
motion symmetry involving the embedding tensor and the top-form
potential. In fact, the coefficient of the $A$ term in
\eqref{covder} can be arbitrarily changed by a redefinition of the
top-form potential in which the 3-form
$C_{\mu\nu\rho}{}^{\cM\cN,\cP}$ is shifted by terms proportional to
$B_{[\mu\nu}{}^{\langle\cM\cN}A_{\rho]}{}^{\cP\rangle}$. In general,
there are several equivalent ways to present the gauge
transformations that are all related  via redefinitions of
fields/parameters and/or adding further equations of motion
symmetries. This will be of relevance when comparing our results
with the ones predicted by $E_{11}$, see the next section. Note that
the equations of motion of $B_{\mu\nu}{}^{\cM\cN}$ and
$C_{\mu\nu\rho}{}^{\cM\cN,\cP}$ give back the constancy of
$\Theta_{\cM\cN}$ and the quadratic constraints.

Using a particular choice of basis we now wish to  determine the
gauge transformations of $B$ and $C$, which are required for the
gauge invariance of the action \eqref{defaction}. (For their
supersymmetry transformations see \cite{deWit:2008ta}.) First of
all, the Chern-Simons term varies as
 \begin{eqnarray}
  \delta_{\Lambda}{\cal L}_{\rm CS}  &=&  -\ft14 g
  \varepsilon^{\mu\nu\rho}D_{\mu}\Theta_{\cM\cN}
  D_{\nu}\Lambda^{\cM}A_{\rho}{}^{\cN} \\ \nonumber
  &&+\ft16 g^2
  \varepsilon^{\mu\nu\rho}\Theta_{\cK\cP}\Theta_{\cL(\cM}f^{\cK\cL}{}_{\cN)}
  A_{\mu}{}^{\cP}A_{\nu}{}^{\cM}D_{\rho}\Lambda^{\cN}\;.
 \end{eqnarray}
Also the scalar-kinetic term is no longer gauge-invariant, since the
$P_{\mu}^{A}$ vary according to
 \begin{eqnarray}
  \delta_{\Lambda} P_{\mu}^{A} &=& gD_{\mu}\Theta_{\cM\cN}
  \Lambda^{\cM}{\cal V}^{\cN A}\;.
 \end{eqnarray}
In addition we have to remember the variation of $A_{\mu}{}^{\cM}$
inside the derivative $D_{\mu}\Theta_{\cM\cN}$. This gives a
contribution proportional to $D\Lambda B$ and the quadratic
constraint and can therefore be canceled by an extra variation of
the top-form. Finally, the T-tensor transforms as
 \bea
  \delta T_{\cA|\cB} \ = \ -2g{\cal Q}_{\cM\cN,\cP}{\cal
  V}^{\cM}{}_{\cA}{\cal V}^{\cN}{}_{\cB}\Lambda^{\cP}\;,
 \eea
and, consequently, the scalar potential varies into the quadratic
constraint. Collecting these terms, the non-invariance of the
Lagrangian can be compensated by introducing the following
transformation rules
 \begin{eqnarray}\label{defvar}
  \delta B_{\mu\nu}{}^{\cM\cN} &=&
  D_{[\mu}\Lambda^{\langle\cM}A_{\nu]}{}^{\cN\rangle}
  -\Lambda^{\langle\cM}\tilde{J}_{\mu\nu}{}^{\cN\rangle}\,,
  \\ \nonumber
  \delta C_{\mu\nu\rho}{}^{\cM\cN,\cP} &=&
  -3D_{[\mu}\Lambda^{\langle\cP}B_{\nu\rho]}{}^{\cM\cN\rangle}
  +A_{[\mu}{}^{\langle\cP}A_{\nu}{}^{\cM}D_{\rho]}\Lambda^{\cN\rangle}
  \\ \nonumber
  &&+\tfrac{1}{16}ge\varepsilon_{\mu\nu\rho}\Lambda^{\langle\cP}\big(-\ft17G^{\cM|\cK|}G^{\cN\rangle\cL}
  -G^{\cM|\cK|}\eta^{\cN\rangle\cL}\big)\Theta_{\cK\cL}
  \;.
 \end{eqnarray}
At this point let us note again that the explicit form of the
projectors indicated in \eqref{defvar} is not required, since in the
variation of the Lagrangian these terms are always multiplied by
$\partial_{\mu}\Theta_{\cM\cN}$ or the quadratic constraint, and so
their projection is manifest.

Next we are going to determine the gauge variations of $B$ and $C$
under their own parameter, $\Lambda_{\mu}$ and $\Lambda_{\mu\nu}$,
respectively. We first consider the gauge transformations with
parameter $\Lambda_\mu$. Defining $\delta
B_{\mu\nu}{}^{\cM\cN}=D_{[\mu}\Lambda_{\nu]}{}^{\cM\cN}$ does not
leave \eqref{defaction} invariant, since the `covariant' derivatives
$D_\mu$ do not commute.\footnote{It turns out that using the
derivative $D_\mu$ in this expression corresponds to a particular
choice of basis for the parameter $\Lambda_{\mu\nu}$.} Rather one
finds the variation
 \begin{eqnarray}\label{extravar}
  \delta \left(\ft14  g \varepsilon^{\mu\nu\rho} D_{\mu}\Theta_{\cM\cN}
  B_{\nu\rho}{}^{\cM\cN}\right)
   &=&
  \ft18 g\varepsilon^{\mu\nu\rho}\Lambda_{\mu}{}^{\cM\cN}
  [D_{\nu},D_{\rho}]\Theta_{\cM\cN} \\ \nonumber
  &= &
  -\ft14
  g^2\varepsilon^{\mu\nu\rho}\Lambda_{\mu}{}^{\cM\cN}F_{\nu\rho}{}^{\cP}\Theta_{\cK\cP}\Theta_{\cL(\cM}f^{\cK\cL}{}_{\cN)}
  \\ \nonumber
  &&+\ft12 g^2 \varepsilon^{\mu\nu\rho}\Lambda_{\mu}{}^{\cM\cN}A_{\nu}{}^{\cP}(D_{\rho}\Theta_{\cK\cP})\Theta_{\cL(\cM}f^{\cK\cL}{}_{\cN)}
  \\ \nonumber
  &&+\ft12 g^3\varepsilon^{\mu\nu\rho}\Theta_{\cP\cQ}\Theta_{\cR(\cK}f^{\cQ\cR}{}_{\cL)}
  \Theta_{\cS\cM}f^{\cS\cL}{}_{\cN}\Lambda_{\mu}{}^{\cM\cN}A_{\nu}{}^{\cP}A_{\rho}{}^{\cK}\;.
 \end{eqnarray}
To compensate these we add a St\" uckelberg like shift
transformation to the gauge vectors,
$\delta^{\prime}A_{\mu}{}^{\cM}=-g\Theta_{\cN(\cK}f^{\cM\cN}{}_{\cL)}\Lambda_{\mu}{}^{\cK\cL}$.
The Chern-Simons term then picks up an additional variation, which
precisely cancels the variation in \eqref{extravar} proportional to
the field strength. Apart from that, the $P_{\mu}^{A}$ vary as
 \bea
  \delta P_{\mu}^{A} \ = \
  g^2\Theta_{\cM\cN}\Theta_{\cP(\cK}f^{\cM\cP}{}_{\cL)}\Lambda_{\mu}{}^{\cK\cL}
  {\cal V}^{\cN A}\;,
 \eea
while the variation of $A_{\mu}{}^{\cM}$ inside the derivative
$D_{\mu}\Theta_{\cM\cN}$ also gives rise to a term proportional to
the quadratic constraint, which both can be absorbed into an extra
transformation of $C$.

We next consider the gauge symmetry of the top-form, $\delta
C_{\mu\nu\rho}{}^{\cM\cN,\cK}=D_{[\mu}\Lambda_{\nu\rho]}{}^{\cM\cN,\cP}$.
The action transforms into a total derivative and terms proportional
to $D_{\mu}\Theta_{\cM\cN}$. The latter can be compensated by a
shift transformation of $B$ under $\Lambda_{\mu\nu}$. This
establishes the gauge-invariance of the action with respect to
$\Lambda_{\mu\nu}$.

Summarizing, we have shown that the bosonic gauge transformations
that leave the action corresponding to the Lagrangian
\eqref{defaction} invariant are given by
 \begin{eqnarray}\label{fullvar}
  \delta A_{\mu}{}^{\cM} &=& D_{\mu}\Lambda^{\cM}-g\Theta_{\cN\cK}
  f^{\cM\cN}{}_{\cL}\Lambda_{\mu}{}^{\cK\cL}\;, \\ \nonumber
  \delta B_{\mu\nu}{}^{\cM\cN} &=& D_{[\mu}\Lambda_{\nu]}{}^{\cM\cN}
  +\delta A_{[\mu}{}^{\langle\cM}A_{\nu]}{}^{\cN\rangle}-\Lambda^{\langle\cM}
  \tilde{J}_{\mu\nu}{}^{\cN\rangle} \\ \nonumber
  &&+\ft23g
  \Theta_{\cK\cL}f^{\cK\langle\cM}{}_{\cP}\big(\Lambda_{\mu\nu}{}^{|\cL\cP|,\cN\rangle}
  -\Lambda_{\mu\nu}{}^{\cN\rangle\cP,\cL}\big)\,, \\ \nonumber
  \delta C_{\mu\nu\rho}{}^{\cM\cN,\cP} &=&
  D_{[\mu}\Lambda_{\nu\rho]}{}^{\cM\cN,\cP}
  -3\hspace{0.15em}\delta A_{[\mu}{}^{\langle\cP}B_{\nu\rho]}{}^{\cM\cN\rangle}
  +A_{[\mu}{}^{\langle\cP}A_{\nu}{}^{\cM}\delta
  A_{\rho]}{}^{\cN\rangle} \\ \nonumber
  &&+  \ft32
  \Lambda_{[\mu}{}^{\langle\cM\cN}\tilde{J}_{\nu\rho]}{}^{\cP\rangle}
  +\tfrac{1}{16}ge\varepsilon_{\mu\nu\rho}\Lambda^{\langle\cP}\big(-\ft17G^{\cM|\cK|}G^{\cN\rangle\cL}
  -G^{\cM|\cK|}\eta^{\cN\rangle\cL}\big)\Theta_{\cK\cL}
  \;.
 \end{eqnarray}

As a consistency check we verify the closure of the gauge algebra.
We first consider the $[\pf 1,\pf 1]$ commutator. Here we indicate
the generators associated to the corresponding $p$--forms with $\pf
1$, $\pf 2$ and $\pf 3$ and their gauge variation with
$\delta^{(p)}$. We find
 \begin{eqnarray}\label{closure}
  \big[\delta_{\Lambda}^{(1)},\delta_{\Sigma}^{(1)}\big]A_{\mu}{}^{\cM}  &=& \big(\delta_{\tilde{\Lambda}}^{(1)}
  +\delta_{\tilde{\Lambda}}^{(2)}+\delta_{\tilde{\Lambda}}^{(3)}\big)A_{\mu}{}^{\cM} \;, \\
  \nonumber
  \big[
  \delta_{\Lambda}^{(1)},\delta_{\Sigma}^{(1)}\big]B_{\mu\nu}{}^{\cM\cN}
  &=& \big(\delta_{\tilde{\Lambda}}^{(1)}
  +\delta_{\tilde{\Lambda}}^{(2)}+\delta_{\tilde{\Lambda}}^{(3)}\big)B_{\mu\nu}{}^{\cM\cN}
  \\ \nonumber
  &&+gf^{\cM\cK}{}_{\cP}\Theta_{\cK\cL}\left(F_{\mu\nu}{}^{\cL}+\tilde{J}_{\mu\nu}{}^{\cL}\right)
  \Lambda^{[\cP}\Sigma^{\cN]}\;, \\ \nonumber
  \big[\delta_{\Lambda}^{(1)},\delta_{\Sigma}^{(1)}\big]C_{\mu\nu\rho}{}^{\cM\cN,\cP}
  &=&\big(\delta_{\tilde{\Lambda}}^{(1)}
  +\delta_{\tilde{\Lambda}}^{(2)}+\delta_{\tilde{\Lambda}}^{(3)}\big)C_{\mu\nu\rho}{}^{\cM\cN,\cP}\;,
 \end{eqnarray}
where the transformation parameters are given by
 \begin{eqnarray}
  {\tilde \Lambda}^{\cM} & = &
  -g\Theta_{\cN\cK}f^{\cM\cN}{}_{\cL}\Lambda^{[\cK}\Sigma^{\cL]}\;,
  \\ \nonumber
  \tilde{\Lambda}_{\mu}{}^{\cM\cN} &=&
  D_{\mu}\Lambda^{\langle\cM}\Sigma^
  {\cN\rangle}-D_{\mu}\Sigma^{\langle\cM}\Lambda^{\cN\rangle}\;, \\ \nonumber
  {\tilde \Lambda}_{\mu\nu}{}^{\cM\cN,\cP} &=&
  3\Lambda^{\langle\cM}\tilde{J}_{\mu\nu}{}^{\cN}\Sigma^{\cP\rangle}\;.
 \end{eqnarray}
We note that in deriving \eqref{closure} we have made use of the
scalar equations of motion, the constancy of the embedding tensor
and the quadratic constraint, i.e.~the closure is only on-shell. For
simplicity, we do not give these terms explicitly in the above
expressions, but just indicate that the on-shell closure on the
deformation potential is guaranteed by the duality relation
\eqref{covdual} between vectors and scalars. One may wonder whether
it is possible to close this algebra off-shell by using the extra
symmetries discussed in \cite{deWit:2008ta} (see eq.~\eqref{extra}).
However, on the deformation potential they act as
$\delta_{\chi}B_{\mu\nu}{}^{\cM\cN}\sim
A_{\mu}{}^{\cM}\delta_{\chi}A_{\nu}{}^{\cN}$ and are therefore not
of the form required by \eqref{closure} --- apart from the fact that
it would still not be clear how to eliminate the other equations of
motion. We conclude that there is no straightforward way to achieve
an off-shell closure, though the possibility of introducing
auxiliary fields, etc., might be worth to investigate.

The only other non-trivial commutator to consider is $[\pf 1, \pf
2]$. We find, for instance,
 \begin{eqnarray}
  \big[\delta_{\Lambda}^{(1)},\delta_{\Sigma}^{(2)}\big]B_{\mu\nu}{}^{\cM\cN}
  &=& \delta_{\tilde{\Sigma}}^{(3)}B_{\mu\nu}{}^{\cM\cN}\;,
 \end{eqnarray}
where
 \bea
  \tilde{\Sigma}_{\mu\nu}{}^{\cM\cN,\cP} \ = \ 3
  \Sigma_{[\mu}{}^{\langle\cM\cN}D_{\nu]}\Lambda^{\cP\rangle}\;.
 \eea
This concludes our discussion of the commutator algebra.

We end this section by considering the duality relation between the
deformation potential and the embedding tensor. Varying the action
corresponding to \eqref{defaction} with respect to $\Theta_{\cM\cN}$
yields the following  `duality relation':
 \bea\label{thetadual}
  e^{-1}\varepsilon^{\mu\nu\rho}G_{\mu\nu\rho}{}^{\cM\cN}
  +2A_{\mu}{}^{\langle\cM}J^{\mu
  \cN\rangle} \ = \
  \tfrac{1}{4}g
  G^{\cM\cN,\cK\cL}\Theta_{\cK\cL}\;.
 \eea
Here we have defined
 \bea\label{gaugedG}
 \begin{split}
  G_{\mu\nu\rho}{}^{\cM\cN} \ = \
  D_{[\mu}B_{\nu\rho]}{}^{\cM\cN}&+A_{[\mu}{}^{\langle\cM}\partial^{}_{\nu}
  A_{\rho]}{}^{\cN\rangle}-2g\Theta_{\cK\cL}f^{\cK\langle\cM}{}_{\cP}A_{[\mu}{}^{\cN\rangle}B_{\nu\rho]}{}^{\cL\cP}
  \\
  &-\ft23 g\Theta_{\cK\cL}f^{\cK\langle\cM}{}_{\cP}\big(C_{\mu\nu\rho}{}^{|\cL\cP|,\cN\rangle}-
  C_{\mu\nu\rho}{}^{\cN\rangle\cP,\cL}-A_{[\mu}{}^{\cN\rangle}A_{\nu}{}^{\cL}A_{\rho]}{}^{\cP}\big)
  \;.
 \end{split}
 \eea
Let us stress that $G$ is not a gauge-covariant field strength. For
instance, ignoring the scalar potential and its variation for the
moment, one finds that the left-hand side of \eqref{thetadual}
varies under $\Lambda^{\cM}$ as
 \begin{eqnarray}
  \delta_{\Lambda}\big(\varepsilon^{\mu\nu\rho}G_{\mu\nu\rho}{}^{\cM\cN}
  +2eA_{\mu}{}^{\langle\cM}J^{\mu\cN\rangle}\big) &=&
  -2\Lambda^{\langle\cM}
  D_{\mu}\big(e J^{\mu\cN\rangle}\big)\\ \nonumber
  &&+g\varepsilon^{\mu\nu\rho}f^{\cK\langle\cM}{}_{\cP}
  A_{\mu}{}^{\cN\rangle}\Theta_{\cK\cL}\big(F_{\nu\rho}{}^{\cL}
  +\tilde{J}_{\nu\rho}{}^{\cL}\big)\Lambda^{\cP}\;,
 \end{eqnarray}
i.e.~it rotates into the scalar equations of motion and the duality
relation. In other words, despite the fact that $G$ does not
transform `covariantly', the entire set of bosonic field equations
is gauge-invariant. This concludes our discussion about
three-dimensional gauged supergravity.

\section{$E_{11}$ and extended ungauged supergravity}
In this section we are going to make the correspondence between
ungauged supergravity and the Kac-Moody algebra $E_{11}$ more
precise. \emph{A priori} there is a puzzle here since the $\Theta=0$
limit of gauged supergravity leads to an ungauged theory in which
the deformation and top-form potentials have disappeared from the
Lagrangian. On the other hand, these same potentials are contained
in the level decomposition of $E_{11}$. In this section we will show
that a specific extended ungauged limit of gauged supergravity
exists whose symmetries on all $p$--form potentials $(p=0,1,2,3)$
are in precise correspondence to the non-linearly realized
symmetries of (a truncation of) $E_{11}$, and which still contains
all forms up to the top-form potentials. In the next subsection we
first discuss the non-linear realization of $E_{11}$. In the
following subsection we will discuss how the same result can be
obtained by taking a limit of gauged supergravity.

\subsection{Non-linear realization of $E_{11}$}

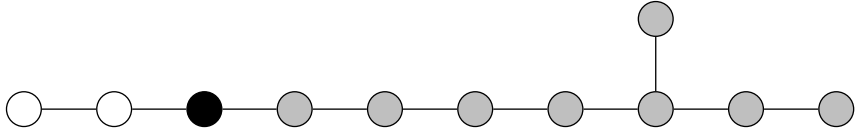
\begin{figure}[t]
\begin{center}
\begin{pspicture}(0,0)(9,1)
\dualityNode{7,1}{N11811119684} \normalNode{0,0}{N21811119684}
\normalNode{1,0}{N31811119684} \disabledNode{2,0}{N41811119684}
\dualityNode{3,0}{N51811119684} \dualityNode{4,0}{N61811119684}
\dualityNode{5,0}{N71811119684} \dualityNode{6,0}{N81811119684}
\dualityNode{7,0}{N91811119684} \dualityNode{8,0}{N101811119684}
\dualityNode{9,0}{N111811119684}
\singleConnection{N21811119684}{N31811119684}
\singleConnection{N31811119684}{N41811119684}
\singleConnection{N41811119684}{N51811119684}
\singleConnection{N51811119684}{N61811119684}
\singleConnection{N61811119684}{N71811119684}
\singleConnection{N71811119684}{N81811119684}
\singleConnection{N81811119684}{N91811119684}
\singleConnection{N91811119684}{N101811119684}
\singleConnection{N101811119684}{N111811119684}
\singleConnection{N11811119684}{N91811119684}
\end{pspicture}
\end{center}
\caption{
 $E_{11}$ decomposed under $SL(3,\R) \times E_{8(8)}$. The white nodes represent
 $SL(3,\R)$, the gray nodes $E_{8(8)}$, and the black node is `disabled'.}
\label{E11_decomp}
\end{figure}


We first consider the non-linear realization of $E_{11}$. In the
case at hand  we have to perform a level decomposition with respect
to $SL(3,\R)\times E_{8(8)}$ (see figure \ref{E11_decomp}), which
are the space-time and duality subgroups. We restrict to the
$p$--form algebra, which means that we truncate to generators that
are totally antisymmetric in their `space-time' indices
$\mu,\nu,\rho$ \cite{Bergshoeff:2007vb}. Specifically, this gives
rise to generators $X^\mu{}_{\cM}$, $Y^{\mu\nu}{}_{\cM\cN}$, and
$Z^{\mu\nu\rho}{}_{\cM\cN,\cP}$ at level 1, 2 and 3, whose
representations are given in table \ref{E11_table}
\cite{Bergshoeff:2007qi}. We note that the level 2 generator is in
precise correspondence with the linear constraint found for gauged
supergravity, while the level 3 generator is consistent with the
quadratic constraint. However, $E_{11}$ allows for an additional
top-form in $\bf{248}$, which is not related to a quadratic
constraint.\footnote{Such top-forms could be related to space-time
filling branes. Similar appearances of extra top-forms have been
encountered in $D=9,10$ \cite{Bergshoeff:2007vb,Bergshoeff:2007qi}.}
Here, these will not be considered further, and by abuse of notation
we will denote the generator in which this additional $\bf{248}$ has
been projected out also by $Z^{\mu\nu\rho}{}_{\cM\cN,\cP}$. The
non-trivial Lie brackets read
 \begin{eqnarray}\label{Liebrackets}
  [X^{\mu}{}_{\cM} , X^{\nu}{}_{\cN}] &=&  2 Y^{\mu\nu}{}_{\cM\cN} \;,
  \\ \nonumber
  [Y^{\mu\nu}{}_{\cM\cN} , X^{\rho}{}_{\cP}] &=&
  3 Z^{\mu\nu\rho}{}_{\cM\cN,\cP} \;.
 \end{eqnarray}

In order to determine the non-linearly realized $E_{11}$ symmetry in
this truncation, we have to introduce a group valued coset
representative,
 \bea\label{cosetrepr}
  \cV \ = \ \exp\left(A_{\mu}{}^{\cM}
  X^{\mu}{}_{\cM}+B_{\mu\nu}{}^{\cM\cN}Y^{\mu\nu}{}_{\cM\cN}+
  C_{\mu\nu\rho}{}^{\cM\cN,\cP}Z^{\mu\nu\rho}{}_{\cM\cN,\cP}\right)\;.
 \eea
Here we have chosen the Borel gauge, in which only positive level
generators enter. The action of the rigid symmetry group is given by
 \bea
  {\cal V} \rightarrow g{\cal V} h^{-1}(x)\;, \qquad  g\in E_{11}\;,
 \eea
where $h(x)$ denotes a local transformation which, if necessary,
restores the chosen gauge for ${\cal V}$. However, after the
gauge-fixing to positive levels in (\ref{cosetrepr}), it is
sufficient for our purpose to consider the symmetry action by a
group element truncated to positive level as well,
 \bea\label{groupelement}
  g \ = \ \exp\left(\Lambda_{\mu}{}^{\cM}X^{\mu}{}_{\cM}+
  \Lambda_{\mu\nu}{}^{\cM\cN}Y^{\mu\nu}{}_{\cM\cN}+
  \Lambda_{\mu\nu\rho}{}^{\cM\cN,\cP}Z^{\mu\nu\rho}{}_{\cM\cN,\cP}\right)\;.
 \eea
Consequently, a compensating local transformation is not required.
Acting with (\ref{groupelement}) on the coset representative
\eqref{cosetrepr}, yields by use of the Baker-Campbell-Hausdorff
formula and the Lie algebra \eqref{Liebrackets} the following global
symmetry transformations
 \begin{eqnarray}\label{e11trans}
  \delta A_{\mu}{}^{\cM} &=&  \Lambda_{\mu}{}^{\cM} \;, \nonumber\\
  \delta B_{\mu\nu}{}^{\cM\cN} &=& \Lambda_{\mu\nu}{}^{\cM\cN}
  +\Lambda_{[\mu}{}^{\langle\cM}A_{\nu]}{}^{\cN\rangle}\;, \\
  \nonumber
  \delta  C_{\mu\nu\rho}{}^{\cM\cN,\cP} &=&
  \Lambda_{\mu\nu\rho}{}^{\cM\cN,\cP}-\ft32B_{[\mu\nu}{}^{\langle\cM\cN}\Lambda_{\rho]}{}^{\cP\rangle}
  +\ft32\Lambda_{\mu\nu}{}^{\langle\cM\cN}A_{\rho}{}^{\cP\rangle}-\ft12A_{[\mu}{}^{\langle\cM}\Lambda_{\nu}{}^{\cN}A_{\rho]}{}^{\cP\rangle}\;.
 \end{eqnarray}
In the next section we will show that these symmetries can also be
obtained in a specific limit of supergravity.

\begin{table}[t]
\begin{centering}
\begin{tabular}{|c|c|c|}
    \hline
    Level   & $SL(3,\R) \times E_{8(8)}$ representation           & Generator \\
    \hline
    \hline
    &  & \\ [-10pt]
    1   & $( {\bf 3}, {\bf 248})$                    & $X^\mu{}_{\cM}$ \\
    &  & \\ [-10pt]
    \hline
    &  & \\ [-10pt]
    2   & $( {\bf \bar 3}, {\bf 1} \oplus {\bf 3875})$    & $Y^{\mu\nu}{}_{\cM\cN}$ \\
    &  & \\ [-10pt]
    \hline
    &  & \\ [-10pt]
    3   & $( {\bf 1}, {\bf 248} \oplus {\bf 3875} \oplus {\bf 147250} )$    & $Z^{\mu\nu\rho}{}_{\cM\cN,\cP}$ \\ [4pt]
    \hline
\end{tabular}
\caption{
    $SL(3,\R) \times E_{8(8)}$ representations within $E_{11}$ up to level 3,
    of which the $SL(3,\R)$ part is totally antisymmetric.}
\label{E11_table}
\end{centering}
\end{table}

\subsection{Extended ungauged supergravity}
In order to see the symmetry \eqref{e11trans} in supergravity one
has to consider a special ungauged limit. More precisely, taking the
standard limit to ungauged supergravity, $g\rightarrow 0$, is
equivalent to setting the embedding tensor to zero. This in turn
eliminates the 1-, 2- and 3-forms from the action and, consequently,
makes the comparison with $E_{11}$  problematic. Moreover, from
\eqref{fullvar} one infers that in this naive limit scalar-dependent
terms survive in the transformation rules as, for instance,
$\delta_{\Lambda}B_{\mu\nu}{}^{\cM\cN}=-\Lambda^{\langle\cM}\tilde{J}_{\mu\nu}{}^{\cN\rangle}$.
These are not predicted by $E_{11}$, and so one has to take a more
subtle limit. To be concrete, we first perform the following
rescaling of the fields,
 \bea\label{rescaling}
  A_{\mu}{}^{\cM} & \rightarrow & g^{1/2} A_{\mu}{}^{\cM}\;, \nonumber \\
  B_{\mu\nu}{}^{\cM\cN} & \rightarrow & g  B_{\mu\nu}{}^{\cM\cN} \;, \\
  C_{\mu\nu\rho}{}^{\cM\cN,\cP} & \rightarrow & g^{3/2} \nonumber
  C_{\mu\nu\rho}{}^{\cM\cN,\cP}\;,
 \eea
and then take the limit $g\rightarrow 0$. This yields the
Lagrangian,
 \bea\label{limit}
  {\cal L} \ = \ {\cal L}_0 -
  \ft14\varepsilon^{\mu\nu\rho}\Theta_{\cM\cN}G^{(0)}_{\mu\nu\rho}{}^{\cM\cN}\;,
 \eea
where ${\cal L}_0$ denotes the standard Lagrangian of ungauged
supergravity. Here, $G^{(0)}_{\mu\nu\rho}{}^{\cM\cN}$ is the
$g\rightarrow 0$ limit of $G_{\mu\nu\rho}{}^{\cM\cN}$, given by
 \bea\label{2-form}
  G^{(0)}_{\mu\nu\rho}{}^{\cM\cN} \ = \
  \partial_{[\mu}B_{\nu\rho]}{}^{\cM\cN}
  +A_{[\mu}{}^{\langle\cM}\partial_{\nu}A_{\rho]}{}^{\cN\rangle}\;.
 \eea
We note that, in contrast to the gauged expression in
\eqref{gaugedG}, this represents a gauge-invariant field strength.
The Lagrangian \eqref{limit} is equivalent to standard ungauged
supergravity in that it merely represents an extension by
topological 1- and 2-forms with vanishing
curvatures.\footnote{Recently, a similar use of topological fields
in the context of the Kac-Moody approach has been made in
\cite{Gomis:2007gb}.} To be more precise, the embedding tensor now
acts as a Lagrange multiplier that sets the curvature of the 2-form
to zero, while the field equations for $A_{\mu}{}^{\cM}$ imply that
their (abelian) field strengths vanish.

Let us now turn to the symmetries that survive in this limit.
Rescaling the symmetry parameters as for the fields in
\eqref{rescaling}, i.e.~$\Lambda^{\cM}\rightarrow
g^{1/2}\Lambda^{\cM}$, etc., yields the following limit of the gauge
symmetries (\ref{fullvar}),
 \begin{eqnarray}
  \delta_{\Lambda}A_{\mu}{}^{\cM} &=& \partial_{\mu}\Lambda^{\cM}\;,
  \\ \nonumber
  \delta_{\Lambda}B_{\mu\nu}{}^{\cM\cN} &=&
  \partial_{[\mu}\Lambda_{\nu]}{}^{\cM\cN}+\partial_{[\mu}\Lambda^{\langle\cM}
  A_{\nu]}{}^{\cN\rangle} \;, \\ \nonumber
  \delta_{\Lambda}\hat{C}_{\mu\nu\rho}{}^{\cM\cN,\cP} &=&
  \partial_{[\mu}\Lambda_{\nu\rho]}{}^{\cM\cN,\cP}
  -\ft32\partial_{[\mu}\Lambda^{\langle\cP}B_{\nu\rho]}{}^{\cM\cN\rangle}
  +\ft32\partial_{[\mu}\Lambda_{\nu}{}^{\langle\cM\cN}A_{\rho]}{}^{\cP\rangle}
  -\ft12A_{[\mu}{}^{\langle\cP}A_{\nu}{}^{\cM}
  \partial_{\rho]}\Lambda^{\cN\rangle}\;.
 \end{eqnarray}
Here we performed the field redefinition
 \bea
  \hat C_{\mu\nu\rho}{}^{\cM\cN,\cP} \ = \
    C_{\mu\nu\rho}{}^{\cM\cN,\cP} + \tfrac 3 2 A_{[\mu}{}^{\langle\cP} B_{\nu\rho]}{}^{\cM\cN\rangle} \; .
 \eea
In particular we observe that the scalar-dependent terms drop out.
Specifying the gauge parameters to linear space-time dependence
according to
 \bea
  \Lambda^{\cM} \ = \ x^{\rho}\Lambda_{\rho}{}^{\cM}\;, \qquad
  \Lambda_{\mu}{}^{\cM\cN} \ = \
  x^{\rho}\Lambda_{\rho\mu}{}^{\cM\cN}\;, \qquad
  \Lambda_{\mu\nu}{}^{\cM\cN,\cP} \ = \
  x^{\rho}\Lambda_{\rho\mu\nu}{}^{\cM\cN,\cP}\;,
 \eea
gives precisely the global symmetry in \eqref{e11trans} predicted by
$E_{11}$.

We note that in the $g\rightarrow 0$ limit the top-form vanishes
from the Lagrangian but does have a well-defined gauge
transformation rule which is in accordance with the $E_{11}$
algebra. Therefore only the (truncated) $E_{10}$ subalgebra is
non-trivially realized at the level of the Lagrangian. Finally, in
the $g\rightarrow 0$ limit, the gauge algebra closes
\textit{off-shell}, as it should be since it matches the $E_{11}$
results, which a priori do not contain information about the
equations of motion.

\section{Conclusions}

In this note we compared a level decomposition based on the very
extended Kac-Moody algebra $E_{11}$ with a particular limit of
maximal three-dimensional gauged supergravity. Before taking the
limit, the gauged supergravity theory contains besides scalars and
vectors also deformation and top-form potentials on which the gauge
algebra, which we determined explicitly, closes \emph{on-shell}.
After taking the limit we are left with a Lagrangian containing
scalars, vectors and deformation potentials on which the gauge
algebra closes \emph{off-shell}. This gauge algebra allows for a
rigid truncation, which in turn realizes an $E_{10}$ subalgebra of
$E_{11}$. To obtain the full $E_{11}$ prediction one must include
the top-form potentials which, however, do not occur in the
Lagrangian.\footnote{This could change if one considers the
inclusion of source terms for spacetime filling branes.} It is
intriguing to note that the lowest-order terms in the variation
$\delta C$ of the top-form as predicted by $E_{11}$ are, from the
supergravity side, required for canceling the \textit{higher-order}
terms in $\Theta$ in the variation of the action. So in this sense,
$E_{11}$ does know about the gauging.

It is natural to expect that the need for a rescaling in order to
match the $E_{11}$ prediction for the deformation and top-form
potentials appears in any dimension. In particular, it would be
interesting to verify this in the case of $D=5$ analyzed in
\cite{Riccioni:2007ni}. However, there the full gauge
transformations have been given up to the 3-forms, for which a
rescaling is not required. Thus, a comparison with our results must
await an exhaustive analysis of the 4- and 5-forms in $D=5$.

Moreover, it would be interesting to extend, for three dimensions,
the relation between extended ungauged supergravity and $E_{10}$
and/or $E_{11}$ to the gauged case. Since, in going from the
ungauged to the gauged case, the closure of the gauge algebra goes
from off-shell to on-shell we expect that dynamics will play a
non-trivial role in this extension. Recently, for the case of
$E_{11}$, a proposal for such a relationship in the gauged case has
been made \cite{Riccioni:2007ni}. It would be interesting to see
wether this proposal yields the details and in particular the
on-shell closure of the three-dimensional gauge algebra. Since
dynamics is involved it would be interesting to also consider the
relationship from the point of view of the $E_{10}$ coset model
\cite{Nicolai:2003fw, Damour:2002cu, Damour:2004zy} where dynamics
is naturally included via the sigma model equations of motion. This
would extend the analysis of \cite{Kleinschmidt:2004dy} for $D=10$
massive supergravity to a case where the gauging of a symmetry is
involved. We hope to report on the results of such an investigation
in the nearby future \cite{inprogress}.

\subsection*{Acknowledgments}
For useful comments and discussions we would like to thank M.~de
Roo, B.~de Wit, A.~Kleinschmidt, H.~Samtleben and M.~Trigiante. This
work was partially supported by the EU MRTN-CT-2004-005104 grant
 and by  the INTAS Project 1000008-7928.

\end{document}